\documentclass{article}
\usepackage{spconf,amsmath,graphicx,booktabs,makecell}

\title{DYNAMIC SPEECH ENDPOINT DETECTION WITH REGRESSION TARGETS}
\name{\begin{tabular}{c}
Dawei Liang$^{\star}$, Hang Su$^{\dagger}$, Tarun Singh$^{\dagger}$, Jay Mahadeokar$^{\dagger}$, Shanil Puri$^{\dagger}$, Jiedan Zhu$^{\dagger}$, \\
Edison Thomaz$^{\star}$, Mike Seltzer$^{\dagger}$
\end{tabular}}
\address{$^{\star}$ University of Texas at Austin, $^{\dagger}$ Meta AI.}

\begin{document}
%
\maketitle
\begin{abstract}
Interactive voice assistants have been widely used as input interfaces in various scenarios, e.g. on smart homes devices, wearables and on AR devices. Detecting the end of a speech query, i.e. speech end-pointing, is an important task for voice assistants to interact with users. Traditionally, speech end-pointing is based on pure classification methods along with arbitrary binary targets. In this paper, we propose a novel regression-based speech end-pointing model, which enables an end-pointer to adjust its detection behavior based on context of user queries. Specifically, we present a pause modeling method and show its effectiveness for dynamic end-pointing. Based on our experiments with vendor-collected smartphone and wearables speech queries, our strategy shows a better trade-off between endpointing latency and accuracy, compared to the traditional classification-based method. We further discuss the benefits of this model and generalization of the framework in the paper.

\end{abstract}
\section{Introduction}
\label{sec:intro}

With rapid development of speech technologies in recent years, interactive voice assistants have been widely adopted as a mainstream for intelligent user interface \cite{porcheron2018voice, pradhan2018accessibility, garg2020he, bentley2018understanding}. By taking users' speech queries, these systems are able to perform a variety of tasks from basic question answering, music playing, calling and messaging, to device control. As an initial step, a voice assistant needs to determine the time point when a user finishes the query so that it knows when to close the microphone and continue downstream processing (e.g. language understanding and taking action). This process is often referred to as speech \textit{endpoint detection} or \textit{end-pointing}. In practice, the challenge for speech end-pointing lies in the conflict goals of fast response and endpoint accuracy (avoiding early cuts of user speech). Specifically, a rapid close of the microphone and response to user queries brings a better user experience if endpointed correctly, but this inevitably increases the risk of early-cut of user queries. The performance of an end-pointing system is thus evaluated on how this conflict is resolved in practical cases. 

Canonically, voice activity detection (VAD) has been widely used for speech end-pointing \cite{shin2000speech, hariharan2001robust}. In the VAD setting, a model is developed to distinguish speech segments from non-speech segments, which usually includes silence, music, and background noises \cite{chengalvarayan1999robust, chang2018temporal, liang2022automated}. The end of the query can then be determined if a fixed duration of silence is observed by the VAD system. However, this approach is not reliable enough as pointed out by later work \cite{ferrer2003prosody, shannon2017improved}. The fact that a VAD system is typically not trained to distinguish pauses within and at the end of queries prevents the system from capturing enough acoustic cues related to end-of-query detection, such as speaking rhythm or filler sounds \cite{shannon2017improved}.



Recent researches on speech end-pointing mostly focus on classification methods, where a dedicated end-pointer is developed to classify audio frames into end-of-query frames and other frames \cite{shannon2017improved, chang2017endpoint}. The success of the Long Short-Term Memory (LSTM) \cite{hochreiter1997long} architecture contributes to this method. Some other efforts include additional text decoding features \cite{maas2018combining} or user personalized i-vectors \cite{jayasimha2021personalizing} to better fit the end-pointer for specific acoustic environments. In an end-to-end automated speech recognition (ASR) system, the end-pointer may also be jointly optimized with the recognition model \cite{chang2019joint}. Despite the promising results, all of the above works focus on binary detection of end-of-query with hard labels (e.g. 0 and 1). In real scenarios, however, an endpointer shall adjust its endpointing aggressiveness based on semantic, prosodic or other speaking patterns in the query. The traditional binary targets for classification can be less flexible in this respect.

In this paper we study a novel speech end-pointing strategy based on regression method. In this setup, an end-pointer is optimized to fit soft-coded targets during training, and the training targets are set by considering expected pause given semantic context of the queries. By testing on both 14.4M smartphone speech queries and 467K wearables user queries, we show that our proposed method effectively reduces the response delay of end-pointing while maintaining a comparable accuracy performance as the conventional classification-based method. 


\section{Regression-based Endpoint Modeling}
\label{sec:regression}

\subsection{Principle}
\begin{figure}[t]

%

\begin{minipage}[t]{1.0\linewidth}
  \centering
  \centerline{\includegraphics[width=8.5cm]{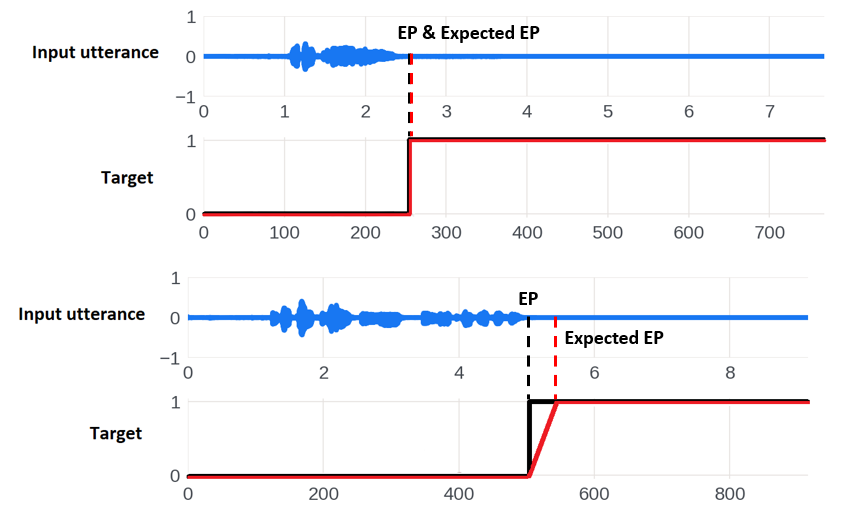}}
\end{minipage}

\caption{Sample targets for end-pointer training based on hard-coded binary values (target in black) and soft-coded float values (target in red). EP: endpoint.} 
\label{target}
\vspace{-8pt}
\end{figure}

In a traditional classification setup, a speech end-pointing model is trained against targets of hard-coded values (0 and 1). At prediction stage, a decision threshold is picked and compared against model's output end-pointing confidence. In a regression setting, the prediction stage remains the same, but the training targets are changed to be continuous float values from 0 to 1, representing the confidence of a frame being part of end-of-query. Specifically, we apply a transition curve to the area between user speech end and the expected endpoint position (i.e. position where we are confident to endpoint) (Figure \ref{target}). The motivation of our regression-based modeling includes: 1. the model should have an increased confidence on endpointing as more silence is observed; 2. the aggressiveness of the endpointing behavior shall be adjustable, which corresponds to the slope of the transition region. As an extreme case, a regression model with 0 transition region would have the same targets as a classification model. The top plot of Figure \ref{target} is an example where the training targets of a regression model and classification model are the same. At the bottom, the regression target enforces the model to be less aggressive, with a smoother slope in the transition region. 

\begin{figure}[t]
  \centering
  \centerline{\includegraphics[width=9cm]{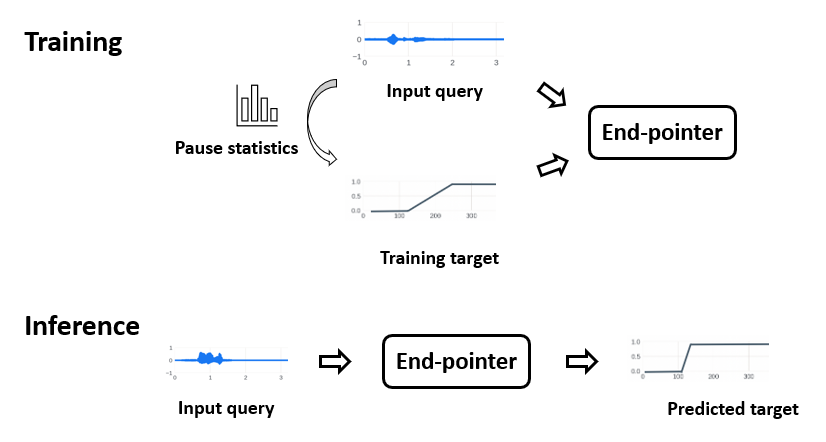}}
  \caption{The overall pipeline of our proposed method.}\medskip

\label{fig:pipeline}
\end{figure}

Figure \ref{fig:pipeline} presents the overall pipeline of end-pointing with regression. To enable a dynamic output behavior of the end-pointer, the transition slope of the training targets can be quantified based on unique patterns (e.g., expected pause) of individual input queries. In this paper, we propose a method to compute and leverage the pause statistics of queries for this purpose (Section \ref{sec:pause}). During inference, the regression model shares the same setup as the classification model, i.e., taking speech features as inputs, and applying a decision threshold to the model outputs for endpoint decision. 

\subsection{Loss Function}
In our study, we use the mean-square-error (MSE) as loss function for end-pointer training. The loss function is defined as below:
\begin{equation}
    l = \frac{\sum^n_{i=1} (\hat{y_i} - y_i)^2}{n}
\end{equation}
where $i$ is the $i$th frame of an utterance in the dataset, $n$ is the total number of frames in the dataset, $\hat{y_i}$ and $y_i$ are the output of the end-pointer and training target, respectively. In our experiments, we found that regression with the binary targets yields the same performance as the traditional classification setup. Furthermore, the MSE loss tends to be more stable for model convergence than the L1 loss.

\section{Speaker Pause Modeling}
\label{sec:pause}

To enable dynamic adjustments of endpointing behavior, we explore the expected pause following a speech segment. Our intuition is that ideal end-pointing aggressiveness should be controlled by the expected pause following a query, and the expected pause is related to semantics. For example, the chance of a speaker continuing  speaking following an utterance \textit{"what's the weather"} is bigger than that following the utterance \textit{"what's the weather in Seattle next week"}. By calculating the expected pause following a speech segment, we can adjust the slope of the transition region to be smoother for the first utterance, so that the end-pointer can learn to wait a bit longer for the first utterance than the second. We hope to estimate the expected duration of each training query and use it to adjust the targets for these queries. 

We model the pause for a given text context as a Gaussian variable:

\begin{equation}
    T_T|C \sim \mathcal{N}(\mu,\,\sigma^{2})
\end{equation}

Here $T_T$ is the pause duration for a text query, $C$ is the text context of the query, $\mu$ and $\sigma$ are parameters for the Gaussian variable. The parameters of this model can be estimated by aggregating pause statistics of utterances with a common context (prefix). For example, the query "\textit{what is the weather}" is considered as a prefix for the queries "\textit{what is the weather}", "\textit{what is the weather in Seattle}", and "\textit{what is the weather today}", and all these queries contribute to the statistics of pause duration given query "\textit{what is the weather}". Note that utterances with transcription "\textit{what is the weather}" have  zero pause, indicating it is a complete utterance itself. In this work, the common prefix definition requires a strict match of transcriptions. Hence, the utterances "\textit{what is the weather}" and "\textit{how is the weather}" are considered to be two independent contexts, though they are semantically similar. Ideally, these utterances with similar semantics should  be grouped together. Also, we used the 95$^{th}$ percentile statistics in our experiment for the expected pause duration estimate rather than mean $\mu$ to reduce the risk of early-cutting.

We further model the pause for a given speech query as follows:

\begin{equation}
    T_S = T_T \cdot R
\end{equation}

Here, $T_S$ is the pause duration for a speech query, $T_T$ is the duration for a text query, and $R$ is the speaking rate. The intuition here is that different queries were spoken at different speaking rates, and the slower a speaker speaks, the longer the expected pause. The speaking rate factor can be estimated using the ratio between the duration of a speech query and the average duration of all those queries with the same prefix (duration of the prefix part only). 


\section{Experimental setup}
\label{sec:experimental}
\subsection{Data}
We used two datasets to evaluate our method - both are collected by 3rd party data vendors. The first one is collected using smartphones whereas the second one is collected using wearable devices (smart glasses). We refer to the two datasets as smartphone data and wearables data in this paper, respectively. The smartphone dataset contains clean user speech that are spoken in fluent manner, with little background noise. In total, 14.4M queries were collected, categorized into 55 domain contexts based on the text transcription. The most common contexts are \textit{music} (1.4M), \textit{weather} (1.3M), and \textit{device handling} (1.2M). The average word count per query is 6. The wearables dataset is collected when users interacts with a voice assistant on smart glasses, with real life noises in the background. A considerable amount of speech pauses and dis-fluency are identified in this data. It contains 447K queries, with an average word count of 11 per query. Both datasets contain speech and ground truth transcriptions. Ground truth speech endpoints are calculated by aligning speech to transcription and measure the last word endtime. From the alignments, we can also collect pause duration (if any) following each word in the queries.

For both datasets, we randomly split the queries with a ratio of 95:5 for model development and testing. A development set was further split out in random during training for learning rate adjustment. The training set and testing set share the same set of speakers.

\begin{table*}[t]%
\begin{center}
\begin{tabular}{cccccc}
  \toprule

  \small \textbf{Threshold} & \small \textbf{Early-cut rate (\%)} & \small \textbf{P50 (ms)} & \small \textbf{P75 (ms)} & \small \textbf{P90 (ms)} & \small \textbf{P99 (ms)}\\
  \midrule
  
    \small 0.50 / 0.63      & \small 3.39 / 3.38 & \small 160 / \textbf{120}& \small 180 / \textbf{150}& \small 230 / \textbf{210} & \small \textbf{480} / 530\\
    
    \small 0.60 / 0.74   & \small 2.45 / 2.38 & \small 170 / \textbf{120}& \small 200 / \textbf{160} & \small 250 / \textbf{240} & \small \textbf{530} / 600\\
    
    \small 0.70 / 0.82 & \small 1.74 / 1.67& \small 180 / \textbf{130} & \small 210 / \textbf{170} & \small 280 / \textbf{260} & \small \textbf{590} / 660\\
  \bottomrule
\end{tabular}
\caption{End-pointing performance with classification / regression-based models for the clean speech dataset.}
\vspace{-12pt}
\label{result1}

\end{center}

\end{table*}%

\begin{table*}[t]%
\begin{center}
\begin{tabular}{cccccc}
  \toprule

  \small \textbf{Threshold} & \small \textbf{Early-cut rate (\%)} & \small \textbf{P50 (ms)} & \small \textbf{P75 (ms)} & \small \textbf{P90 (ms)} & \small \textbf{P99 (ms)}\\
  \midrule
    
    \small 0.56 / 0.50      & \small 14.83 / 14.81 & \small 420 / \textbf{350}& \small 590 / \textbf{450}& \small 860 / \textbf{810} & \small \textbf{1990} / 2500\\
        
    \small 0.60 / 0.59   & \small 12.82 / 12.75 & \small 470 / \textbf{430}& \small 670 / \textbf{510} & \small 970 / \textbf{870} & \small \textbf{2050} / 2500\\
    
    \small 0.70 / 0.67 & \small 10.53 / 10.37& \small 580 / \textbf{500} & \small 780 / \textbf{650} & \small \textbf{1100} / 1130 & \small \textbf{2120} / 2500\\
        
  \bottomrule
\end{tabular}
\caption{End-pointing performance with classification / regression-based models for the noisy speech dataset.}
\vspace{-12pt}
\label{result2}

\end{center}

\end{table*}%

\subsection{Model Configuration}
Our end-pointer is an LSTM neural network inspired by prior work \cite{shannon2017improved, maas2018combining}. Specifically, the model consists of three uni-directional LSTM layers with a hidden unit size of 128 per layer. For the baseline classification model, the output is a fully-connected layer of two neurons, transformed by Log-Softmax activation. For the regression model, the output contains only a single neuron with sigmoid activation. During training, the baseline model leveraged the regular cross-entropy loss, while the regression model uses MSE loss. The learning rate was $2\times10^{-4}$ / $2\times10^{-3}$ for each model, and reduced by a factor of 0.5 if no improvement was observed on the validation set. The Adam optimizer \cite{kingma2014adam} was used with a Beta of (0.9, 0.999). The mini batch size was 128. Besides, a sliding window of 10 ms was applied to the outputs of the baseline model at prediction stage for score smoothing. Both models were trained for 15 epochs when learning rate drops to a small enough value. All development was done using PyTorch toolkit\cite{paszke2019pytorch}.

We use 40-dim filter-bank features as inputs to the network. We extracted the features based on a 25 ms window and a stride of 10 ms. The sampling rate of the audio was 16 kHz.

\section{Results}
\label{sec:result}
\subsection{Overall Results}
In total, we obtained 106K and 18K unique prefixes respectively for the smartphone and the wearable training sets. Following the prior work \cite{shannon2017improved, maas2018combining}, we applied the early-cut rate as the model accuracy metric which reports the proportion of early-cut queries out of the entire test population. We do not use WER as a good endpointing metric, because we think it is not linearly correlated to endpointing accuracy, and thus not linearly correlated to user experience on endpointing. Specifically, an early cut in the begining of an utterance and at the end of an utterance are both early cuts and both cause users to repeat their queries, but their WER differs greatly. For the latency measure, we used the $50^{th}$, $75^{th}$, $90^{th}$, and $99^{th}$ percentile statistics of latency values out of the predicted results, noted as P50, P75,  P90, and P99, respectively. For both metrics, a smaller value is more desirable. Tables \ref{result1} and \ref{result2} summarize the overall model performance of our study, based on sample decision thresholds from around 0.5. Given the closest early-cut rate for each pair of comparisons, we can see that the prediction latency measured by P50, P75, and P90 tends to be improved by the regression model on both datasets. This demonstrates a better balance of end-pointing behavior achieved by the regression model. For P99, the classification model was consistently better. This is in fact reasonable, because P99 shows the very tail latency of prediction, and the regression model was enforced to have a very smooth performance for queries of potentially a high expectation of succeeding pause. For the typical user experience of latency, regression-based end-pointing is still more desirable, as demonstrated by P50 and P75.

\subsection{Discussions}
\begin{figure}[t]

%
\begin{minipage}[b]{1.0\linewidth}
  \centering
  \centerline{\includegraphics[width=8cm]{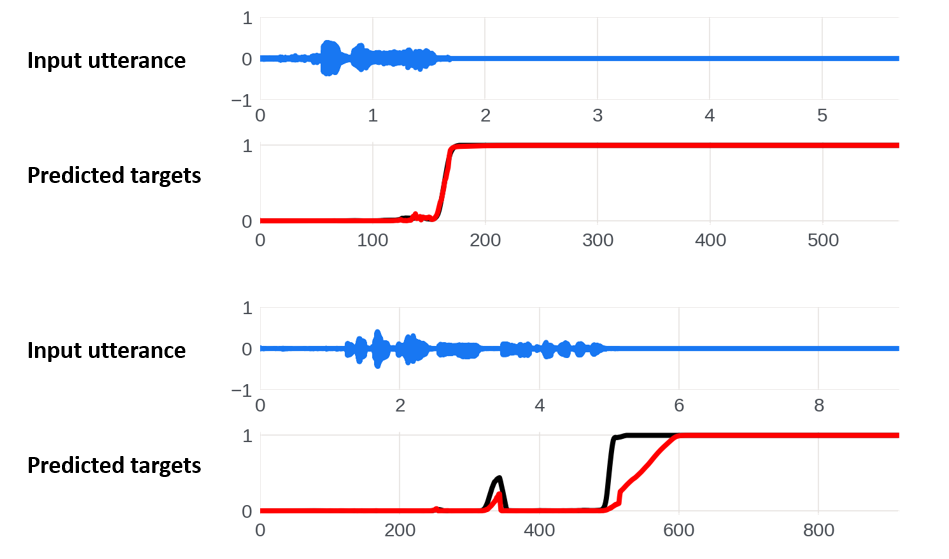}}
\end{minipage}

\caption{Comparison of end-pointer outputs based on classification (black) and regression (red). The training targets of the regression model were adjusted by max pause within queries.}
\label{vis}
\end{figure}

To further study the potential of generalization for regression-based end-pointing, we visualized the model outputs based on a simpler setting. In this setting, the training target slope of the end-pointer was adjusted by hand-crafted criteria based on the max pause duration within a query (0: 10 ms slope duration; $>$ 90 ms: 1,000 ms slope duration). Figure \ref{vis} shows predictions of the regression model for two sample queries. It can be seen that the regression model can achieve the same output behavior as classification-based end-pointing for a query with little within-sentence pauses (top plot). However, the output of the regression model is smoother for a query with significant within-sentence pauses (bottom plot). The plots help to demonstrate that regression-based end-pointing can be applied to different target adjustment strategies.

In addition, we noticed that regression-based end-pointing can help to maintain the model stability for end-pointers of lower complexity. In an extra study, we halved the hidden units of the LSTM layers for our neural end-pointer. By testing with the noisy data, the early-cut rate of the regression model increased to 18.77\% at a threshold of 0.5. The P50, P75, P90 and P99 values also increased to 380 ms, 550 ms, 970 ms, and 2,370 ms, respectively. However, at the closest early-cut rate (27.95\%, threshold = 0.81), the P50, P75, P90 and P99 results of the classification model significantly increased to 1,770 ms, 2,260 ms, 2,480 ms, and 2,690 ms, respectively. A possible explanation is that the abstraction of the unique prefixes from the original queries helps to generalize the end-pointer on the patterns of the queries. This benefit is meaningful for end-pointing on the edge, especially for devices with limited computational capabilities.


%
%
%

\section{Conclusion and Future Work}
\label{sec:conclusions}

In this paper, we proposed a novel regression-based speech end-pointing model, which points to a new direction for solving the early cut issues in endpointing without sacrificing latency. We then describe how we utilize this model for dynamic end-pointing by adjusting training targets based on expected pause of speech queries. Based on experiments with speech data collected by smartphones and wearables, the proposed regression model shows a better performance on latency-accuracy trade-off, with an improved P50, P75 and P90, specifically, while maintaining a comparable early-cut performance to the traditional method. This regression-based endpointing approach is quite flexible and can be used to incorporate speaking behavior modeling in the future. A better pause model can be explored by grouping utterance into semantics rather than pure prefixes as well.

\vfill\pagebreak

\bibliographystyle{IEEEbib}
\bibliography{refs}

\end{document}